\documentclass[twoside]{article}
\usepackage{fleqn,espcrc2}

\usepackage{graphicx}
\usepackage[figuresright]{rotating}

\newcommand{\AmS}{{\protect\the\textfont2
  A\kern-.1667em\lower.5ex\hbox{M}\kern-.125emS}}

\input epsf
\input colordvi  
\long \def \blockcomment #1\endcomment{}
\def\slashchar#1{\setbox0=\hbox{$#1$}           
   \dimen0=\wd0                                 
   \setbox1=\hbox{/} \dimen1=\wd1               
   \ifdim\dimen0>\dimen1                        
      \rlap{\hbox to \dimen0{\hfil/\hfil}}      
      #1                                        
   \else                                        
      \rlap{\hbox to \dimen1{\hfil$#1$\hfil}}   
      /                                         
   \fi}                                         %
 
\title{Lattice gauge theory: A retrospective}
\author{Michael Creutz \address{Physics Department, Brookhaven
National Laboratory, PO Box 5000, Upton, NY 11973-5000, USA\\
creutz@bnl.gov } 
\thanks{This manuscript has been authored under
contract number DE-AC02-98CH10886 with the U.S.~Department of Energy.
Accordingly, the U.S. Government retains a non-exclusive, royalty-free
license to publish or reproduce the published form of this
contribution, or allow others to do so, for U.S.~Government purposes.}}
       
\begin{document}

\begin{abstract}
I discuss some of the historical circumstances that drove us to use
the lattice as a non-perturbative regulator.  This approach has had
immense success, convincingly demonstrating quark confinement and
obtaining crucial properties of the strong interactions from first
principles.  I wrap up with some challenges for the future.
\vspace{1pc}
\end{abstract}

\maketitle

\section{Introduction}

I am honored to have this opportunity to give the final talk at this
wonderful meeting.  I was originally asked to talk about ``Lattice
gauge theory: past, present, and future.''  Well, I really don't know
much about the future, and if I did, I would be out trying to develop
the ideas.  As for the present, well you have for the last few days
been hearing all about it, and there is too much for me to usefully
review anyway.  Thus the majority of this talk will refer back to the
early days.  I will try to recall the conditions that drove us to this
peculiar approach to regularizing quantum field theory.

\section {Particle physics in the late 60's}

I begin by summarizing the situation in particle physics in the late
60's, when I was a graduate student.  Quantum-electrodynamics had
already been immensely successful, but the basic theory was in some
sense ``done.''  While hard calculations remained, and indeed still
remain, there was no major conceptual advance remaining.

These were the years when the ``eightfold way'' for describing
multiplets of particles had recently gained widespread acceptance.
The idea of ``quarks'' was around, but with considerable caution about
assigning them any physical reality; maybe they were nothing but a
useful mathematical construct.  A few insightful theorists were
working on the weak interactions, and the basic electroweak
unification was beginning to emerge.  The SLAC experiments were
observing substantial inelastic electron-proton scattering at large
angles, and this was interpreted as evidence for substructure, with
the term ``parton'' coming into play.  While occasionally there were
speculations relating quarks and partons, people tended to be rather
cautious about pushing this too hard.

A crucial feature at the time was that the extension of quantum
electrodynamics to a meson-nucleon field theory was failing miserably.
The analog of the electromagnetic coupling had a value about 15, in
comparison with the 1/137 of QED.  This meant that higher order
corrections to perturbative processes were substantially larger than
the initial calculations.  There was no known small parameter in which
to expand.

In frustration over this situation, much of the particle theory
community set aside traditional quantum field theoretical methods and
explored the possibility that particle interactions might be
completely determined by fundamental postulates such as analyticity
and unitarity.  This ``S-matrix'' approach raised the deep question of
just ``what is elementary.''  A delta meson might be regarded as a
combination of a proton and a pion, but it would be just as correct to
regard the proton as a bound state of a pion with a delta.
Furthermore, the particles were all bound together by exchanging
themselves.  These ``dual'' views of the basic objects of the theory
persist today in string theory.

\section {The early 70's}

As we entered the 1970's, partons were increasingly identified with
quarks.  This shift was pushed by two dramatic theoretical
accomplishments.  First was the proof of renormalizability for
non-Abelian gauge theories \cite{renormalizability}, giving confidence
that these elegant mathematical structures \cite{ym} might have
something to do with reality.  Second was the discovery of asymptotic
freedom, the fact that interactions in non-Abelian theories become
weaker at short distances \cite{asymptoticfreedom}.  Indeed, this was
quickly connected with the pointlike structures hinted at in the SLAC
experiments.  Out of these ideas evolved QCD, the theory of quark
confining dynamics.

The viability of this picture depended upon the concept of
``confinement.''  While there was strong evidence for quark
substructure, no free quarks were ever observed.  This was
particularly puzzling given the nearly free nature of their apparent
interactions inside the nucleon.  This returns us to the question of
just ``what is elementary.''  Are the physical particles we see in the
laboratory the fundamental objects or are they these postulated quarks
and gluons?

\begin{figure}
\epsfxsize .8\hsize
\centerline{\epsffile {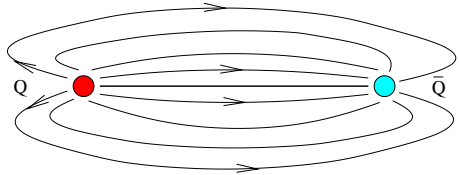}}
\caption {A tube of gluonic flux connects quarks and antiquarks.  The
strength of this string is 14 tons.}
\label{fluxtube}
\end{figure}

Struggling with this paradox led to the now standard flux tube picture
of confinement.  The gluonic fields are analogues of photons except
that they carry ``charge'' with respect to eachother.  Massless
charged particles are rather singular objects, leading to a
conjectured instability that removes zero mass gluons from the
spectrum, but does not violate Gauss's law.  A Coulombic $1/r^2$ field
is a solution of the equations of a massless field, but, without
massless particles, such a spreading of the gluonic flux is not
allowed.  The field lines from a quark cannot end, nor can they spread
in the inverse square law manner.  Instead, as in
Fig.~{\ref{fluxtube}}, the flux lines cluster together, forming a tube
of flux emanating from the quark and ultimately ending on an
antiquark.  This tube is a real physical object, and grows in length
as the quark and antiquark are pulled apart.  The resulting force is
constant at long distance, and can be determined from the slope of the
famous ``Regge trajectories.''  In physical units, it has a strength
of about 14 tons.

The reason a quark cannot be isolated is similar to the reason that a
piece of string cannot have just one end.  Of course one can't have a
piece of string with three ends either, but this is the reason for the
underlying $SU(3)$ group theory.  The confinement phenomenon cannot be
seen in perturbation theory; when the coupling is turned off, the
spectrum becomes free quarks and gluons, dramatically different than
the pions and protons of the interacting theory.

\section{The mid 70's revolution}

We then entered a particularly exciting time for particle physics,
with a series of dramatic events revolutionizing the field.  First was
the discovery of the $J/\psi$ particle \cite{jpsi}.  The
interpretation of this object and its partners as bound states of
heavy quarks provided the hydrogen atom of QCD.  The idea of quarks
became inescapable; field theory was reborn.  The $SU(3)$ non-Abelian
gauge theory of the strong interactions combined with the electroweak
theory became the durable ``standard model.''

This same period also witnessed several remarkable realizations on a
more theoretical front.  Non-linear behaviors in various classical
field theories were shown to have deep consequences for their quantum
counterparts.  Classical ``lumps'' represented a new way to get
particles out of a quantum field theory \cite{lumps}.  Much of the
progress here was in two dimensions, where techniques such as
``bosonization'' showed equivalences between theories of drastically
different appearance.  A boson in one approach might appear as a bound
state of fermions in another, but in terms of the respective
Lagrangian approaches, they were equally fundamental.  Again, we were
faced with the question ``what is elementary?''  Of course modern
string theory is discovering multitudes of ``dualities'' that continue
to raise this same question.

The discovery of such phenomena had deep implications: field theory
can have much more structure than seen from a Feynman diagram analysis
alone.  But this in turn had crucial consequences for practical
calculations.  Field theory is notorious for divergences requiring
regularization.  The bare mass and charge entering the theory are
infinite quantites.  They are not the physical observables, which must
be defined in terms of physical processes.  To carry out calculations,
a ``regulator'' is required to tame the divergences, and when physical
quantities are related to each other, any regulator dependence should
drop out.

The need for controlling infinities had, of course, been known since
the early days of QED.  But all regulators in common use were based on
Feynman diagrams; the theorist would calculate diagrams until one
diverged, and that diagram was then cut off.  Numerous schemes were
devised for this purpose, ranging from the Pauli-Villars approach to
forest formulae to dimensional regularization.  But with the
increasing realization that non-perturbative phenomena were crucial,
it was becoming clear that we needed a ``non-perturbative'' regulator,
independent of diagrams.

\section{The lattice}

The necessary tool appeared through Wilson's lattice theory.  Wilson
presented this as an example of a model with confinement.  The strong
coupling expansion had a finite radius of convergence, allowing a
rigorous demonstration of confinement, albeit in an unphysical limit.
The resulting spectrum had exactly the desired properties; only gauge
singlet bound states of quarks and gluons could move.

This was not the first time that the basic structure of lattice gauge
theory had been written down.  A few years earlier, Wegner
\cite{wegner} presented a $Z_2$ lattice gauge model as an example of a
system posessing a phase transistion but not exhibiting any local
order parameter.  In his thesis, Jan Smit \cite{smit} described using
a lattice regulator to formulate gauge theories outside of
perturbation theory.  The time was clearly ripe for the development of
such a regulator.  Very quickly after Wilson's suggestion, Balian,
Drouffe, and Itzykson \cite{bdi} explored an amazingly wide variety of
aspects of these models.

To reiterate, the primary role of the lattice is to provide a
non-perturbative cutoff.  Space is not really meant to be a crystal,
the lattice is a mathematical trick.  It provides a minimum wavelength
through the lattice spacing $a$, {\it i.e.} a maximum momentum of
$\pi/a$.  Path summations become well defined ordinary integrals.  By
avoiding the convergence difficulties of perturbation theory, the
lattice provides a rigorous route to the definition of quantum field
theory.

The approach, however, had a marvelous side effect.  By discreetly
making the system discrete, it becomes sufficiently well defined to be
placed on a computer.  This was fairly straightforward, and came at
the same time that computers were rapidly growing in power.  Indeed,
numerical simulations and computer capabilities have continued to grow
together, making these efforts the mainstay of lattice gauge theory.

\section{What is a gauge theory?}

The Wilson theory very naturally extends the concept of a gauge theory
to the lattice.  To expand on this I digress into a few of the
``definitions'' of a gauge theory.  Indeed, there are many, and the
lattice approach is closely tied to most.

At the simplest level, a gauge theory is nothing but a generalization
of electromagnetism to include an internal symmetry, i.e. the gauge
fields are given an ``isospin-like'' quantum number.  Both on the
lattice and in the continuum the internal symmetry of the strong gauge
field is $SU(3)$.  Through this extension the gluons acquire charges
with respect to each other.  Infrared difficulties with massless
charged particles lie at the root of the confinement phenomenon.

A common interpretation of a gauge theory is as a dynamics with an
exact local symmetry.  The action should be invariant under a gauge
transformation, such as the familiar
$$
A \longrightarrow g^\dagger A g +ie g^\dagger \partial g
$$
This definition is appealing in that it directly extends to gravity,
where the basic equations are invariant under local coordinate
changes.  This symmetry also immediately applies exactly to the
plaquette form of the Wilson lattice gauge action.  The only
difference between electromagnetism and the strong interactions is
that the fundamental variables are generalized from a simple $U(1)$
for photons to the internal $SU(3)$ symmetry group with eight gluons.

Another point of view considers a gauge theory directly as a theory of
phases.  The interaction of an electron with the electromagnetic field
appears through its wave function acquiring a phase
$$
\psi \longrightarrow \exp (ie\int_{x}^{y} A^\mu dx_\mu) \psi
$$
where the gauge field is integrated along the path traversed by the
particle.  In the rest frame of the particle, this is simply the
statement that the oscillation frequency, or energy, is increased by
the vector potential, {\it i.e.} $E \rightarrow E+eA_0 $.  A
particularly nice thing about this interpretation is that it easily
generalizes to non-Abelian theories; one just replaces the word
``phase'' with ``matrix'' taken from the appropriate group.  For the
strong interations this is the group $SU(3)$, where the 3 comes from
the requirement of 3 quarks making a proton.

Wilson's formulation directly implements a theory of phases on the
lattice.  The basic variables are phases associated with the links of
a lattice.  One such variable is associated with every link connecting
a nearest neighbor pair.  For the strong interactions they are
elements of $SU(3)$.  The bond variables are directed, reversing the
order of two neighbors gives the inverse matrix.

For completeness let me mention one elegant definition of a gauge
theory that does not translate particularly well onto the lattice.
This concept was pushed some time ago by Weinberg \cite{weinberg}, and
looks for interactions in quantum field theory where one is forced to
involve fields that do not transform simply under Lorentz
transformations.  Instead, when one does such a transformation, one
must allow for the possibility of a gauge change.  This point of view
also applies to gravity, where the local formulation requires
introducing Christoffel symbols, the analogue of the gauge potentials.
That these ideas do not formulate well on the lattice is a consequence
of the strong breaking of Lorentz invariance by the lattice regulator.

\begin{figure}
\epsfxsize .6\hsize
\centerline{\epsffile {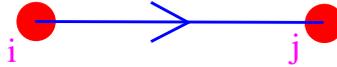}}
\caption {The fundamental variables of lattice gauge theory are
matrices associated with nearest neighbor links.  For the strong
interactions these are $SU(3)$ matrices.}
\end{figure}

\section{Parameters}

The way the link matrices are combined to form the Wilson action is
well known to this audience, so I won't pursue it here.  I do,
however, want to reiterate one of the most remarkable aspects of this
theory, the paucity of adjustable parameters.  To begin with, the
lattice spacing itself is not an observable.  We are using the lattice
to define the theory, and thus for physics we are interested in the
continuum limit $a\rightarrow 0$.  Then there is the coupling
constant, which is also not a physical parameter due to the phenomenon
of asymptotic freedom.  The lattice works directly with a bare
coupling, and in the continuum limit this should vanish
$$
e_0^2 \rightarrow 0
$$ 
In the process, the coupling is replaced by an overall scale for this
vanishing.  Coleman and Weinberg \cite{colemanweinberg} gave this
phenomenon the marvelous name ``dimensional transmutation.''  Of
course an overall scale is not really something we should expect to
calculate from first principles.  Its value would depend on the units
chosen, be they furlongs or light-fortnights.
 
Next consider the quark masses.  Indeed, measured in units of the
asymptotic freedom scale, these are the only free parameters in the
strong interactions.  Their origin remains one of the outstanding
mysteries of particle physics.  The massless limit gives a rather
remarkable theory, one with no undetermined dimensionless parameters.
This limit is not terribly far from reality; chiral symmetry breaking
should give massless pions, and experimentally the pion is
considerably lighter than the next non-strange hadron, the rho.  A
theory of two massless quarks is a fair approximation to the strong
interactions at intermediate energies.  In this limit all
dimensionless ratios should be calculable from first principles,
including quantities such as the rho to nucleon mass ratio.

\begin{figure}
\epsfclipon
\centerline{
\epsfxsize .9\hsize
\epsffile {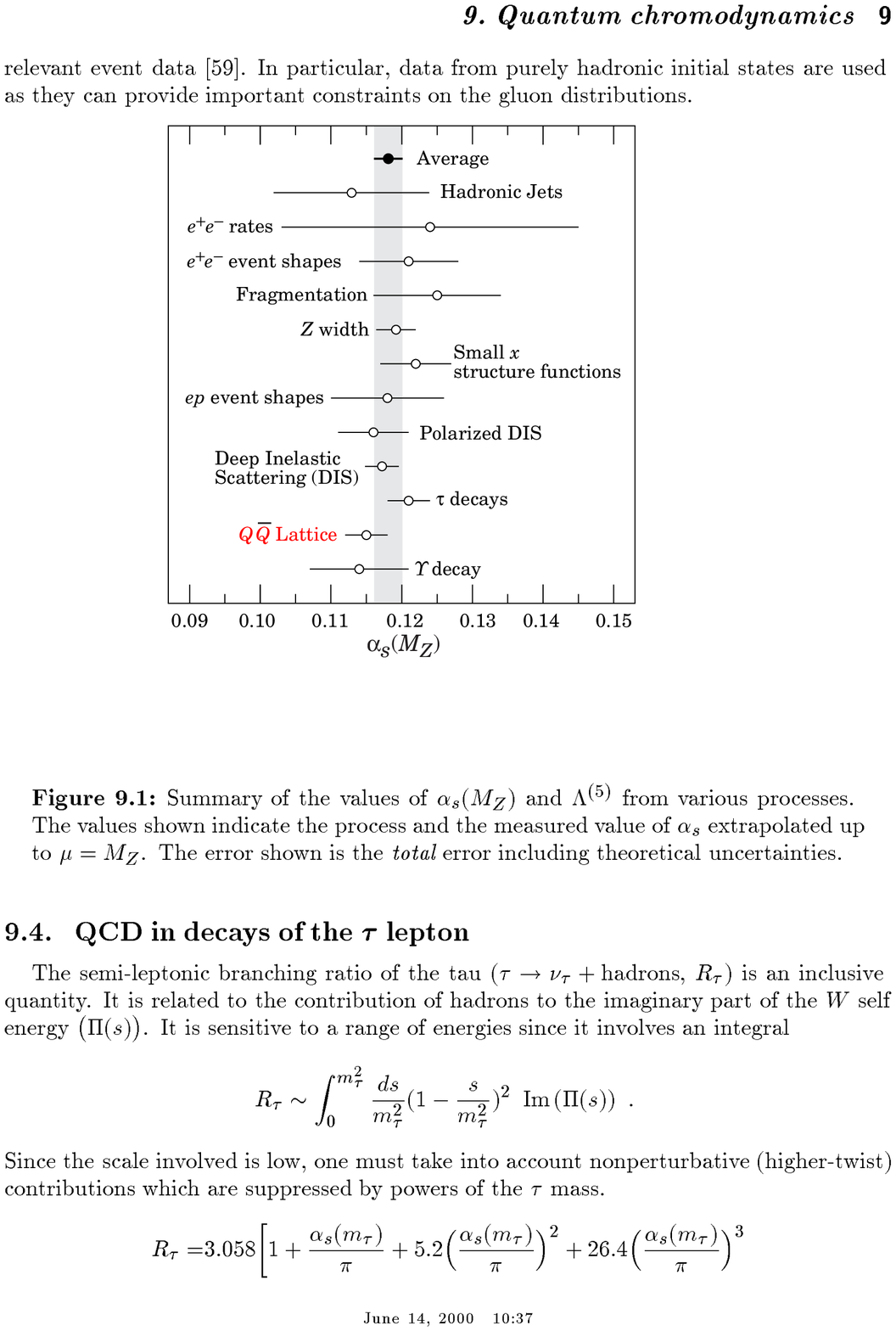}}
\epsfclipoff
\caption{The strong coupling constant is not a parameter, but can be
determined by lattice calculations.  This figure is from the Particle
Data Group summary tables \cite{pdg}.}
\label{alphastrong}
\end{figure}

The strong coupling at any physical scale is not an input parameter,
but should be determined.  Such a calculation has gotten lattice gauge
theory into the famous particle data group tables; see
Fig.\ref{alphastrong}.  With appropriate definition the current result
is\cite{pdg}
$$
\alpha_s(M_Z)=0.115\pm 0.003
$$
where the input is details of the charmonium spectrum.

\section{Numerical simulation}

Of course, as this audience is intimately familiar, large scale
numerical simulations have come to dominate the field.  They are based
on attempts to evaluate the path integral
$$
Z=\int dU e^{-\beta S}
\label{pathintegral}
$$
A direct evaluation of such an integral has pitfalls.  At first sight,
the basic size of the calculation is overwhelming.  Considering a
$10^4$ lattice, small by today standards, there are 40,000 links.  For
each is an $SU(3)$ matrix, parametrized by 8 numbers.  Thus we have a
$10^4\times 4 \times 8 = 320,000$ dimensional integral.  One might try
to replace this with a discrete sum over values of the integrand.  If
we make the extreme approximation of using only two points per
dimension, this gives a sum with
$$
2^{320,000}=3.8\times 10^{96,329}
$$
terms!  Of course, computers are getting pretty fast, but one should
remember that the age of universe is only $\sim 10^{27}$ nanoseconds.

These huge numbers suggest a statistical treatment.  Indeed, the
integral in Eq.~\ref{pathintegral} is formally just a partition
function.  Consider a more familiar statistical system, such as a
glass of Kingfisher.  There are a huge number of ways of arranging the
atoms of carbon, hydrogen, oxygen, etc.~that still leaves us with a
glass of Kingfisher.  We don't need to know all those arrangements, we
only need a dozen or so ``typical'' glasses to know all the important
properties.

This is the basis of the Monte Carlo approach.  The analogy with a
partition function and the role of ${1\over \beta}$ as a temperature
enables the use of standard techniques to obtain ``typical''
equilibrium configurations, where the probability of any given
configuration is given by the Boltzmann weight
$$
P(C)\sim e^{-\beta S(C)}
$$
For this we use a Markov process, making changes in the current
configuration
$$
C\rightarrow C^\prime \rightarrow \ldots
$$
biased by the desired weight.

\begin{figure}
\epsfxsize\hsize
\centerline{\epsffile {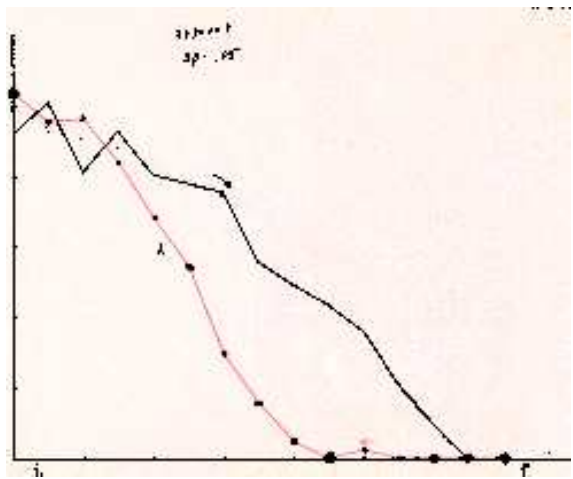}}
\caption{The results of an early lattice gauge simulation of $Z_2$
lattice gauge theory on a $3^4$ lattice.  A rapid thermal cycle shows
hints of hysteresis.  The ``blotches'' are the consequence of not
raising the pen from the paper during a sweep.  The program was
written in basic and run on a programable Hewlett-Packard calculator.}
\label{z2lgt}
\end{figure}

The idea is easily demonstrated with the example of $Z_2$ lattice
gauge theory \cite{creutzjacobsrebbi}.  For this toy model the links
are allowed to take only two values, either plus or minus unity.
One sets up a loop over the lattice variables.  When looking at a
particular link, calculate the probability for it to have value $1$
$$
P(1)={e^{-\beta S(1)}\over e^{-\beta S(1)}+e^{-\beta S(-1)}}
$$ 
Then pull out a roulette wheel and select either 1 or $-1$ biased by
this weight.  Lattice gauge Monte-Carlo programs are by nature quite
simple.  They are basically a set of nested loops surrounding a random
change of the fundamental variables.  Fig.~\ref{z2lgt} shows the
results of my first simulation of this model.

\section{Selected accomplishments}

\begin{figure}
\epsfxsize .9\hsize
\centerline{\epsffile {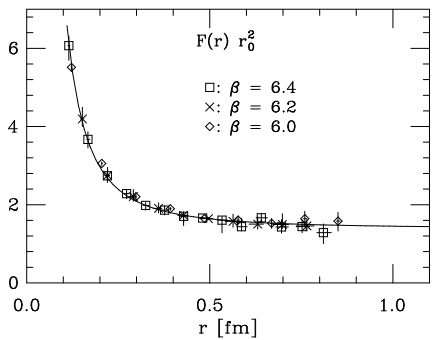}}
\caption{The force between two quarks does not fall to zero as the
distance increases.  This is the confinement phenomenon. (From
Ref.~\cite{michael}).}
\label{force}
\end{figure}

Of course this entire meeting is about the accomplishments of the
technique.  The results have been fantastic, giving first principles
calculations of interacting quantum field theory.  I will just mention
two examples.  The early result that bolstered the lattice into
mainstream particle physics was the convincing demonstration of the
confinement phenomenon.  The force between two quark sources indeed
remains constant at large distances.  A summary of this result is
shown in Fig.~\ref{force}, taken from Ref.~\cite{michael}.

\begin{figure}
\epsfclipon
\epsfxsize .9\hsize
\centerline{\epsffile {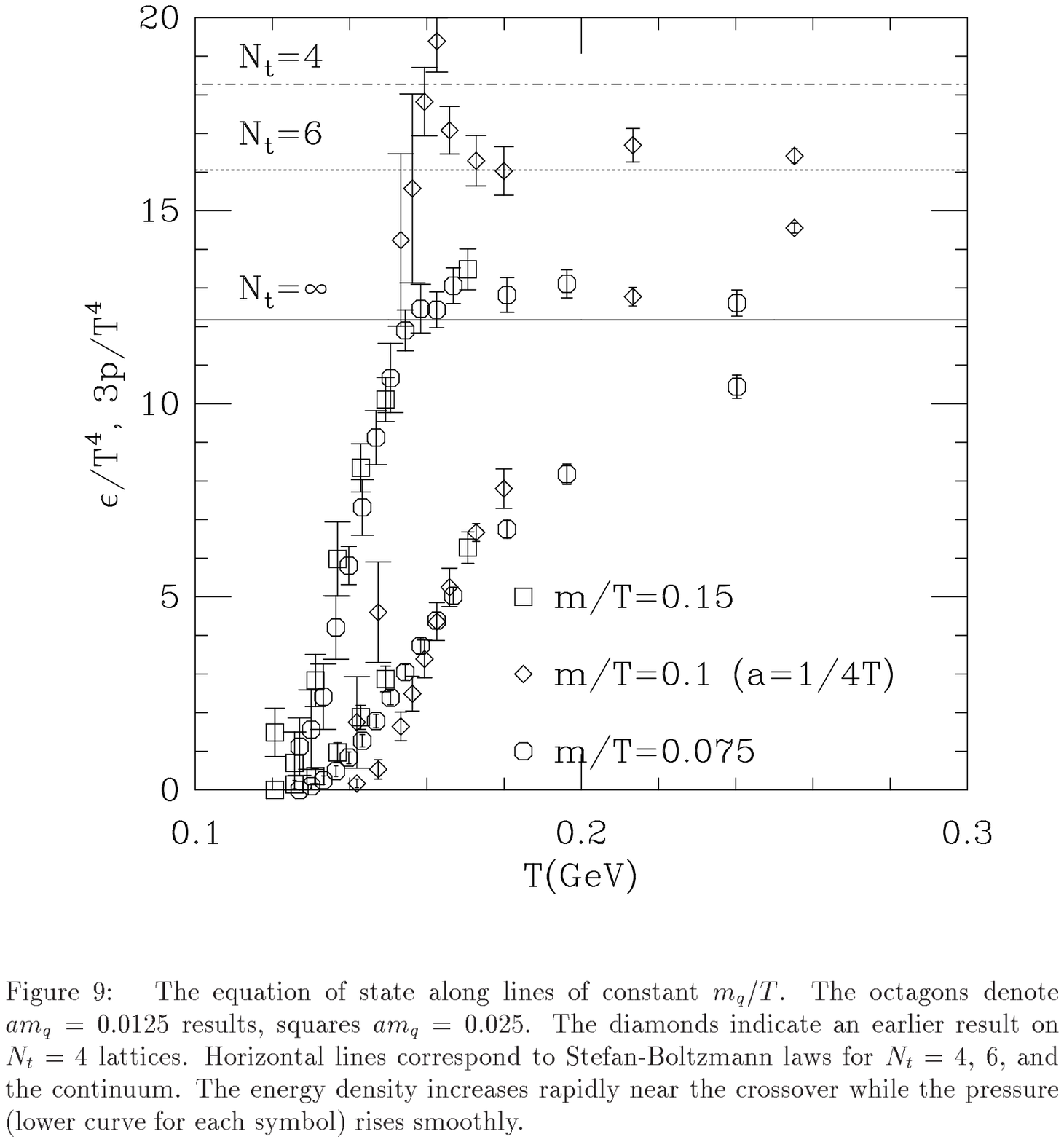}}
\epsfclipoff
\caption{The energy and pressure of the {\ae}ther show a dramatic
structure at a temperature of about 170--190 MeV.  The lattice is a unique
theoretical tool for the study of this transition to a quark-gluon
plasma (From Ref.~\cite{eos}).}
\label{finitetemp}
\end{figure}

Another accomplishment for which the lattice excels over all other
methods has been the study the deconfinement of quarks and gluons into
a plasma at a temperature of about 170--190 Mev\cite{plasma}.  Indeed,
the lattice is a unique quantitative tool capable of making precise
predictions for this temperature.  The method is based on the fact
that the Euclidean path integral in a finite temporal box directly
gives the physical finite temperature partition function, where the
size of the box is proportional to the inverse temperature.  This
transition represents a loss of confining flux tubes in a background
plasma.  Fig.~\ref{finitetemp} shows a recent calculation of this
transition \cite{eos}.

\section{Quarks}

While the gauge sector of lattice gauge theory is in good shape, from
the earliest days fermionic fields have caused annoying difficulties.
Actually there are several apparently unrelated fermion problems.  The
first is an algorithmic one.  The quark operators are not ordinary
numbers, but anticommuting operators in a Grassmann space.  As such
the exponentiated action itself is not a number but rather an
operator.  This makes comparison with random numbers problematic.

Over the years various clever tricks for dealing with this problem
have been developed; we have seen numerous large scale Monte Carlo
simulations involving dynamical fermions.  The algorithms used are all
essentially based on an initial analytic integration of the quarks to
give a determinant.  This, however, is the determinant of a rather
large matrix, the size being the number of lattice sites times the
number of fermion field components, with the latter including spinor,
flavor, and color factors.  In my opinion, the algorithms working
directly with these large matrices remain quite awkward.  I often
wonder if there is some more direct way to treat fermions without the
initial analytic integration.

The algorithmic problem becomes considerably more serious when a
chemical potential generating a background baryon density is present.
In this case the required determinant is not positive; it cannot be
incorporated as a weight in a Monte Carlo procedure.  This is
particularly frustrating in the light of striking predictions of
superconducting phases at large chemical potential
\cite{superconduct}.  This is perhaps the most serious unsolved
problem in lattice gauge theory.

The other fermion problems concern chiral issues.  There are a variety
of reasons that such symmetries are important in physics.  First is
the light nature of the pion, which is traditionally related to the
spontaneous breaking of a chiral symmetry expected to become exact as
the quark masses go to zero.  Second, the standard model itself is
chiral, with the weak bosons coupling to chiral currents.  Third, the
idea of chiral symmetry is frequently used in the development of
unified models as a tool to prevent the generation of large masses and
thus avoid fine tuning.

Despite its importance, chiral symmetry and the lattice have never fit
particularly well together.  I regard this as evidence that the
lattice is trying to tell us something deep.  Indeed, the lattice
fully regulates the theory, and thus all the famous anomalies must be
incorporated explicitly.  It is well known that the standard model is
anomalous if either the quarks or leptons are left out, and this
feature must appear in any valid formulation.

\begin{figure}
\epsfxsize .9\hsize
\centerline{\epsffile {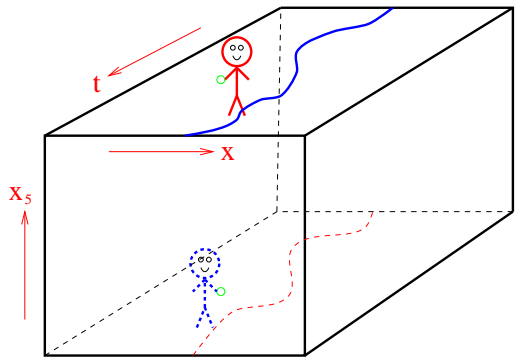}}
\caption{The domain-wall formulation of chiral symmetry regards our
four dimensional world as an interface in an underlying five dimensional
theory}
\label{kaplanfig}
\end{figure}

These issues are currently a topic with lots of activity
\cite{myreview}.  Several schemes for making chiral symmetry more
manifest have been developed, with my current favorite being the
domain-wall formulation, where, as sketched in Fig.~\ref{kaplanfig}
our four dimensional world is an interface in an underlying five
dimensional theory.

\section{My Pet Problems}

As I said at the beginning, I do not know what areas will dominate
lattice gauge theory in the future.  Thus let me conclude by
mentioning some problems that particularly interest me.  These are all
directly connected with the problems of quarks discussed in the
previous section.

The first is the chiral symmetry problem, alluded to above.  Here the
recent developments have put parity conserving theories, such as the
strong interactions, into quite good shape.  The various schemes,
including domain-wall fermions, the overlap formula, and variants on
the Ginsparg-Wilson relation, all quite elegantly give the desired
chiral properties.  Chiral gauge theories themselves, such as the weak
interactions, are not yet completely resolved, but the above
techniques appear to be tantilizingly close to giving a well defined
lattice regularization.  It is still unclear whether the lattice
regularization can simultaneously be fully finite, gauge invariant,
and local.  I expect these issues to be a major topic of continuing
research and look forward to the final resolution.

Chiral symmetry should be expected to have a considerably broader
impact on particle physics in the future.  The problems encountered
are closely related to similar issues with supersymmetry, another area
that does not naturally fit on the lattice.  This also ties in with
the explosive activity in string theory and a possible regularization
of gravity.

The other area in particular need of advancement lies in dynamical
fermion methods.  As I said earlier, I regard all existing algorithms
as frustratingly awkward.  This, plus the fact that sign problem with
a background density remains completely unsolved, suggests that new
ideas are needed.  It has long bothered me that we treat fermions and
bosons so differently in numerical simulations.  Indeed, why is it
that we have to treat them separately?

\end{document}